\documentclass[conference]{IEEEtran}

\IEEEoverridecommandlockouts

\usepackage[utf8]{inputenc}
\usepackage[T1]{fontenc}
\usepackage{times}          

\usepackage{newtxtext,newtxmath}
\usepackage{hyperref}             
\usepackage{microtype}
\usepackage[dvipsnames]{xcolor}   
\usepackage{url}
\usepackage{soul}                 

\usepackage{graphicx}
\usepackage{subcaption}
\usepackage[font=small,labelfont=bf]{caption}
\usepackage{wrapfig}

\usepackage{amsmath}
\usepackage{ntheorem}

\usepackage{amssymb}

\usepackage{nicefrac}
\usepackage{hyperxmp}
\usepackage{booktabs}
\usepackage{multirow}
\usepackage{algorithm}
\usepackage{algorithmic}
\usepackage[most]{tcolorbox}

\usepackage{xspace}
\usepackage{enumitem}
\usepackage{pifont}
\usepackage{minted}       
\usepackage{lineno}       

\usepackage{listings}
\usepackage{cite}
\usepackage{fancyvrb}

\definecolor{BLUE}{named}{Blue}   

\renewcommand{\paragraph}[1]{\vskip 0.05in \noindent {\bf #1.}}

\newcommand{\dataset}{\textit{Defects4C}\xspace}   
\newcommand{\datac}{\textit{Defects4C\_{bgcommit}}\xspace} 
\newcommand{\datab}{\textit{Defects4C\_{bug}}\xspace} 
\newcommand{\datav}{\textit{Defects4C\_{vul}}\xspace}

\newcommand{\datasetcommitsize}{\textbf{9M}\xspace} 
\newcommand{\datasetsize}{\textbf{350}\xspace} 
\newcommand{\datasetsizenorm}{\textbf{248}\xspace} 
\newcommand{\datasetsizecve}{\textbf{102}\xspace}

\newcommand{\datasetsizenormi}{248\xspace}  
\newcommand{\datasetsizecvei}{102\xspace}  

\newcommand{\llmsize}{24\xspace}  

\newcommand{\datasetprojects}{\textbf{41}\xspace} 

\newcommand{\xmark}{\ding{55}}%

\definecolor{gree}{rgb}{0.2,0.5,0}
\definecolor{BLUE}{named}{Blue}

\begin{document}

\title{\dataset: Benchmarking Large Language Model Repair Capability with C/C++ Bugs}


\author{
  \IEEEauthorblockN{%
    Jian Wang\textsuperscript{1,4},\,
    Xiaofei Xie\textsuperscript{1},\,
    Qiang Hu\textsuperscript{2}\IEEEauthorrefmark{2},\,
    Shangqing Liu\textsuperscript{3}\IEEEauthorrefmark{2}\thanks{\IEEEauthorrefmark{2} Co-Corresponding authors: qianghu@tju.edu.cn; shangqingliu@nju.edu.cn},\,
    Jiongchi Yu\textsuperscript{1},\,
    Jiaolong Kong\textsuperscript{1},\,
    and Yi Li\textsuperscript{4}
  }
  \IEEEauthorblockA{%
    \textsuperscript{1}Singapore Management University, Singapore\\
    \textsuperscript{2}Tianjin University, China\\
    \textsuperscript{3}State Key Laboratory for Novel Software Technology, Nanjing University, China\\
    \textsuperscript{4}Nanyang Technological University, Singapore \\
  }
  }


\maketitle

\begin{abstract}
Automated Program Repair (APR) plays a critical role in enhancing the quality and reliability of software systems. While substantial progress has been made in Java-based APR, largely facilitated by benchmarks like Defects4J, there remains a significant gap in research on C/C++ program repair, despite the widespread use of C/C++ and the prevalence of associated vulnerabilities. This gap is primarily due to the lack of high-quality, open-source benchmarks tailored for C/C++.

To address this issue, we introduce \dataset, a comprehensive and executable benchmark specifically designed for C/C++ program repair. Our dataset is constructed from real-world C/C++ repositories and includes a large collection of bug-relevant commits (\datasetcommitsize in total), \datasetsizenorm high-quality buggy functions, and \datasetsizecve vulnerable functions, all paired with test cases for reproduction. These resources enable rigorous evaluation of repair techniques and support the retraining of learning-based approaches for enhanced performance.

Using \dataset, we conduct a comprehensive empirical study evaluating the effectiveness of \llmsize state-of-the-art large language models (LLMs) in repairing C/C++ faults. Our findings offer valuable insights into the strengths and limitations of current LLM-based APR techniques in this domain, highlighting both the need for more robust methods and the critical role of \dataset in advancing future research.

\end{abstract}

\section{Introduction}
Software bugs pose significant security and reliability threats to modern software systems. In safety-critical and large-scale software, even a single defect can lead to severe consequences such as data breaches and system crashes. Fixing such bugs is often challenging and costly, debugging and maintenance activities can account for up to 50\% of the total software development cost, much of which involves time-consuming manual effort for fault localization, root cause analysis, and patch implementation~\cite{britton2013reversible}. 
Given these challenges, automating repair of software bugs has become a crucial research direction. Over the past decade, this area has gained significant traction in both academia and industry, with numerous repair techniques~\cite{xia2023automated,guo2024exploring} proposed to increase software developer productivity and reduce the debugging costs. Moreover, the advent of large language models (LLMs) has demonstrated significant improvements over traditional repair methods, offering superior performance in program repair tasks~\cite{xia2023keep}.

Despite the extensive research on Automated Program Repair (APR), the vast majority of existing work has primarily focused on languages such as Java and Python. This focus is largely driven by the availability of mature and well-structured benchmarks, such as Defects4J~\cite{just2014defects4j} for Java and BugsInPy~\cite{widyasari2020bugsinpy} for Python. These benchmarks provide standardized, reproducible settings for evaluating APR techniques and have played a crucial role in advancing the field.

However, C and C++ continue to serve as the foundation for high-performance and system-level software, powering critical infrastructures such as operating systems, embedded devices, network services, and safety-critical applications. Notably, C/C++ remains the language with the highest number of reported vulnerabilities, accounting for over 50\% of all disclosed open-source vulnerabilities since 2019, according to recent reports~\cite{Secure}. In fact, the annual count of vulnerabilities in C significantly exceeds that of any other programming language. Despite this, C/C++ program repair remains relatively underexplored, and it is still unclear how well existing APR techniques perform when applied to real-world C/C++ bugs and vulnerabilities. A major bottleneck is the absence of a comprehensive, high-quality benchmark dataset, similar to Defects4J, which supports realistic, executable, and testable repair scenarios in C/C++ environments.

While there have been efforts to construct C/C++ defect benchmarks for APR evaluation~\cite{orvalho2022c:C-Pack-IPAs,tan2017codeflaws:CodeFlaw,bohme2017bug:DBGBench,yi2017feasibility,le2015manybugs:ManyBugs,defect4c_ase_demo:BUGSC,gupta2017deepfix:DeepFix}, limitations remain in terms of bug diversity, dataset usability, and scale—all of which are critical for meaningful APR research. For example, benchmarks like DeepFix~\cite{gupta2017deepfix:DeepFix} and Code4Bench~\cite{majd2019code4bench:Code4Bench} derive bugs from student assignments or competitive programming platforms, resulting in simplified buggy functions that do not reflect the complexity of real-world applications. Other benchmarks such as DBGBench~\cite{bohme2017bug:DBGBench} include data from only two projects, which limits their representativeness across software ecosystems. Meanwhile, ManyBugs~\cite{le2015manybugs:ManyBugs} and Prophet~\cite{long2016automatic:Prophet} focus on specific C standards (e.g., C99/C11) and suffer from limited usability, requiring lengthy compilation processes and lacking user-friendly interfaces, which makes them difficult to use in large-scale evaluations~\cite{lutellier2020coconut}. The most recent benchmark, BUG-C++~\cite{defect4c_ase_demo:BUGSC}, collects defect data from GitHub commits; however, we found that some of these commits may not correspond to actual bugs.

Therefore, there remains a pressing need for a high-quality C/C++ bug benchmark that satisfies the key criteria of practicality, diversity, fidelity, and usability, to enable rigorous evaluation and foster the advancement of APR techniques for C/C++ programs.

At the same time, automated program repair techniques have evolved significantly with the emergence of large language models (LLMs). Recent advances in code understanding and generation have demonstrated the remarkable capabilities of LLMs, particularly on Java and Python datasets~\cite{chen2021codex:humaneval,evalplus,xia2023keep}. Recent studies show that LLM-based APR techniques often surpass traditional methods in both bug-fixing accuracy and efficiency~\cite{program_repair_website}. However, these developments have primarily focused on high-level languages, and the effectiveness of LLMs in repairing C/C++ bugs remains largely underexplored, largely due to the absence of a suitable benchmark.

This gap hampers a comprehensive understanding of LLM capabilities and limitations in the context of C/C++ program repair, which poses distinct challenges such as low-level memory manipulation, undefined behavior, and complex control flows. Given the prevalence of bugs and security vulnerabilities in C/C++ software, it is essential to evaluate LLM-based repair techniques on realistic C/C++ faults to uncover their true potential and identify areas for improvement, thereby driving future research and innovation in this critical domain.

To address the aforementioned challenges and gaps, we introduce a new high-quality C/C++ fault benchmark, referred to as \dataset, which comprises two major components: bug-relevant commits (\datac) and curated buggy functions, further categorized into general bugs (\datab) and vulnerabilities (\datav). The \datac dataset includes a broad collection of commit-level changes that are potentially bug-related, making it well-suited for training or fine-tuning data-driven models, despite the possible presence of false positives. In contrast, the buggy function datasets (\datab and \datav) are carefully verified by human experts to ensure correctness and quality, making them ideal for rigorous evaluation of program repair techniques. This design balances the need for large-scale, diverse training data with the requirement for reliable and precise benchmarks for assessment.

Specifically, we first leveraged BigQuery to extract a large number of buggy commits (\textit{40M}) from over \textit{110K} widely used GitHub C/C++ repositories using a set of predefined bug-related keywords. We then filtered the commits based on availability (resulting in \textbf{9M} bug-related commits) and whether the changes were isolated to a single function (leading to \textbf{76K} single-function buggy commits). A unit test matching method was applied to identify corresponding test cases for each buggy function, leaving representative \textbf{ 3,785} buggy commits collected from the top 100 projects with paired tests. To ensure the quality of the dataset for evaluation, we implemented a three-stage human annotation process conducted by three security experts. This process was crucial for eliminating false positives, i.e., cases where commit messages contain bug-related keywords, but the code changes do not actually address bugs or security issues. Our rigorous approach resulted in \datasetsizenorm confirmed bugs (\datab) along with their corresponding unit tests, allowing for bug reproduction and repair validation. 

In addition, we expanded the diversity of the dataset by including a vulnerability dataset (\datav). We first extracted C/C++-related Common Vulnerabilities and Exposures (CVEs) from a publicly available database~\cite{cve_from}. To isolate vulnerable functions, we selected CVEs that provided patched commit IDs, allowing us to retrieve the associated vulnerable and patched functions from the commits. We then applied the unit test matching process to identify corresponding test cases for each vulnerability, ultimately yielding \datasetsizecve vulnerabilities with corresponding unit tests.


To understand the effectiveness of state-of-the-art LLM-based APR techniques in fixing C/C++ bugs or vulnerabilities, we conducted an empirical study using our \dataset benchmark. The study focuses on evaluating the performance of LLM-based APR techniques, incorporating state-of-the-art LLMs. These models are evaluated in
single-round and conversation-based program repair scenarios with various experimental settings. 
Our findings reveal a significant performance gap in LLM-based APRs when addressing C/C++ faults compared to their success with the Defects4J benchmark~(Java). This discrepancy highlights the need for APR techniques specifically tailored for C/C++ fault repair. We further explored the effectiveness of fine-tuning in C/C++ program repair, and while the results show some promise, they remain below acceptable levels. Moreover, a deeper analysis shows that bugs span multiple lines and bugs that require external information to fix in \dataset, are difficult to repair with LLMs, posing a potential direction to propose new fine-tuning methods to handle C/C++ bugs. 

To sum up, we make the following contributions:
\begin{itemize}[leftmargin=*]
 \item  We have developed and publicly released an {executable} C/C++ defect benchmark namely \dataset, comprising \textbf{9M} bug-relevant commits (\datac), \datasetsizenorm buggy functions (\datab) and \datasetsizecve vulnerable functions (\datav), sourced from GitHub open-source projects. It is accessible at the website\footnote{\url{https://sites.google.com/view/defects4c}}. A user-friendly command line interface for ease of use accompanies each sample in this dataset.
\item We conduct a large-scale empirical study focused on assessing the capability of LLM-based APR techniques in repairing C/C++ programs, and exploring the failure patterns made by these techniques. We select state-of-the-art LLMs with various settings for a comprehensive evaluation. Our findings highlight a significant gap and limitations in the current LLMs when fixing C/C++ bugs, especially in contrast to their performance on Java bugs. These results underscore the need for further research and development of C/C++-specific repair techniques and the importance of \dataset.
\end{itemize}

\section{Motivation and Related Work}

\begin{table*}[!t]
\centering
\caption{Existing C/C++ benchmarks for program repair.}
\label{tbl-existingdataset}
\resizebox{.95\linewidth}{!}{
\begin{tabular}{lccc|lccc}
\toprule 
 Dataset & Defects & Projects & Source & Dataset & Defects & Projects & Source\tabularnewline
\midrule 
CodeHunt~\cite{tillmann2014code:CodeHunt} & 195K & N/A & Interview/Contest & ITSP~\cite{ITSP} & 661 & N/A & Assignment\tabularnewline
Code4Bench~\cite{majd2019code4bench:Code4Bench} & 25K & N/A & Interview/Contest & C-Pack-IPAs~\cite{orvalho2022c:C-Pack-IPAs} & 513 & N/A & Assignment\tabularnewline
Prutor/SARD~\cite{das2016prutor:SARD} & 23K & N/A & Interview/Contest & Bugs-C++~\cite{defect4c_ase_demo:BUGSC} & 209 & 22 & Real-World\tabularnewline
SPoC~\cite{kulal2019spoc:SPoC} & 18K & N/A & Interview/Contest & ManyBugs~\cite{le2015manybugs:ManyBugs} & 185 & 9 & Real-World\tabularnewline
CodeFlaws~\cite{tan2017codeflaws:CodeFlaw} & 3.9K & N/A & Interview/Contest & Prophet~\cite{long2016automatic:Prophet} & 69 & 8 & Real-World\tabularnewline
DeepFix~\cite{gupta2017deepfix:DeepFix} & 6.9K & N/A & Assignment & DBGBench~\cite{bohme2017bug:DBGBench} & 27 & 2 & Real-World\tabularnewline
IntroClass~\cite{le2015manybugs:ManyBugs} & 998 & N/A & Assignment & \dataset & \datasetsize & \datasetprojects & Real-World\tabularnewline
\bottomrule 
\end{tabular}
}
\end{table*}




\noindent \textbf{Program Repair.}
Automated Program Repair (APR) techniques aim to generate candidate patches based on the original code and identified buggy locations. Each synthesized patch is subsequently validated against a test suite. Patches that pass all test cases are deemed plausible, whereas those that effectively resolve the underlying bug are considered correct. In general, APR approaches can be categorized into two paradigms: traditional and learning-based methods.

Traditional tools can be broadly divided into three main categories: heuristic-based~\cite{le2016history, le2011genprog, wen2018context}, constraint-based~\cite{le2017s3, long2015staged, mechtaev2016angelix} and template-based~\cite{liu2019avatar, liu2019tbar, martinez2016astor}. However, these methods have some limitations. For example, template-based tools have achieved state-of-the-art performance among traditional methods due to their best repair success rates, but their effectiveness is constrained by a strong reliance on manually crafted templates or domain-specific fix patterns, which limits their generalizability across diverse types of software bugs.

Unlike conventional methods, learning-based approaches can automatically capture semantic relations among parallel bug-fixing pairs. This capability enables the creation of patch solutions that are not only more effective but also contextually aware. There has been a growing focus on learning-based approaches, such as CURE~\cite{jiang2021cure}, RewardRepair~\cite{ye2022neural}, Recoder~\cite{zhu2021syntax}, CoCoNut~\cite{lutellier2020coconut} SelfAPR~\cite{ye2022selfapr} and ITER~\cite{ye2024iter}, which convert APR to Neural Machine Translation (NMT) problem and have shown remarkable potential for enhancing bug repair performance.
Nevertheless, the quality and quantity of the training datasets largely determine the performance of the model. 

Recently, large language models have exhibited powerful capabilities to repair program defects~\cite{jiang2023impact,prenner2022can,sobania2023analysis,xia2023automated}, they mainly focus on the buggy code and treat bug repair as a one-step process, overlooking the interactive and collaborative aspects inherent in bug resolution. Compared to single-round repair, conversation-based program repair techniques~\cite{xia2023conversational,xia2023keep} are proposed to further improve repair performance. These techniques target interaction with LLMs by feeding error messages as input to guide LLMs in generating more accurate output. However, these LLM-based techniques are mainly evaluated on Defects4J~\cite{just2014defects4j}, and it is not clear their effectiveness on C/C++ projects. 

\noindent \textbf{Existing C/C++ Defect Benchmark.}
Table~\ref{tbl-existingdataset} provides a summary of existing C/C++ benchmarks for program repair, including our proposed dataset, \dataset. To date, prevailing benchmarks for C/C++ programs have mostly centred on student programming assignments such as DeepFix~\cite{gupta2017deepfix:DeepFix}, C-Pack-IPAs~\cite{orvalho2022c:C-Pack-IPAs}, and IntroClass~\cite{le2015manybugs:ManyBugs} or online contests such as Code4Bench~\cite{majd2019code4bench:Code4Bench}, CodeHunt~\cite{tillmann2014code:CodeHunt}, Prutor/SARD~\cite{das2016prutor:SARD}, SPoC~\cite{kulal2019spoc:SPoC}, and CodeFlaw~\cite{tan2017codeflaws:CodeFlaw}. As the data source is from assignments or contests, they are relatively impractical in real-world program repair. To construct a more practical benchmark, several works propose to collect programs from real-world projects such as ManyBugs~\cite{le2015manybugs:ManyBugs}, Prophet~\cite{long2016automatic:Prophet}, DBGBench~\cite{bohme2017bug:DBGBench}, and BUG-C++~\cite{defect4c_ase_demo:BUGSC}. These benchmarks also suffer from various limitations. For instance, ManyBugs and Prophet offer low usability and only support outdated versions of C/C++ programs. DBGBench is limited in diversity, as it is collected from only two GitHub projects. BUG-C++ mainly relies on bug-related keywords from commit messages, which contain some that may not be real bugs.

\textbf{Motivation.} In summary, LLM-based methods have shown significant improvement in program repair, particularly on benchmarks like Defects4J. To explore their generalizability, we conducted preliminary experiments on existing C/C++ benchmarks. As shown in Table~\ref{tab:compare}, LLMs perform well on these benchmarks, which often feature simplified, interview- or contest-style programs. However, when applied to real-world C/C++ projects (e.g., those in our dataset), their performance drops substantially. This observation motivates two goals: (1) to construct a realistic benchmark based on real-world C/C++ projects, and (2) to conduct an empirical study on the effectiveness of LLMs in repairing real-world C/C++ bugs.

\begin{table}[t]
  \centering
\caption{Repair performance (Pass@1) on existing benchmarks vs. real-world C/C++ projects.}
  \resizebox{0.99\linewidth}{!}{
    \setlength{\tabcolsep}{1mm}
    \small
    \begin{tabular}{lc|ccc}
      \toprule
      Benchmark (C/C++)      & Source                            & GPT-3.5-Turbo & GPT-4   & CodeLlama-34b-Inst. \\
      \midrule
      DebugBench~\cite{tian2024debugbench}         & Interview/Contest (LeetCode)      & 59.0       & 74.6 & 16.4           \\
      CodeFlaws~\cite{tan2017codeflaws:CodeFlaw}          & Interview/Contest (Codeforces)    & 94.0       & 93.0 & 91.0           \\
      \dataset (ours)    & Real-World                        &  8.5       &   9.0 &  4.0           \\
      \bottomrule
    \end{tabular}
  }
  \label{tab:compare}
\end{table}

\section{Benchmark Construction}

\begin{figure*}[!t]
     \centering
     \vspace{-4mm}
     \includegraphics[scale=0.5,height=5cm]{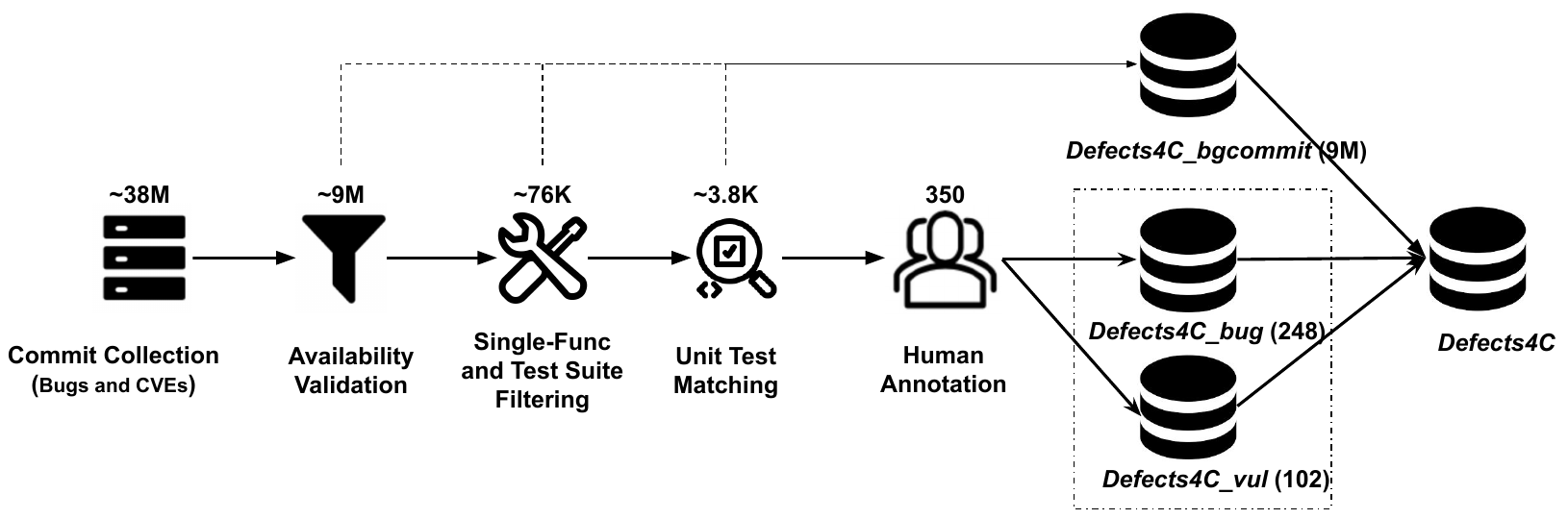}
     \caption{The pipeline of data collection and processing.}
     \label{fig:pipeline}
\end{figure*}

Figure~\ref{fig:pipeline} illustrates the overall workflow of our dataset construction, encompassing raw data collection, test case identification, and human validation. Specifically, we begin by collecting bug-related and vulnerability-related commits from GitHub and the CVE repository. We then apply a series of filtering steps based on repository availability, whether the commit affects a single function, and the presence of test cases.
Next, we develop a test case matching algorithm to identify the specific unit test(s) that validate each fix, filtering out commits that lack a corresponding test case. For the remaining samples, we conduct a rigorous human verification process to confirm the correctness and relevance of the bug fixes and associated tests.
The resulting benchmark, \dataset, is organized into two main components: (1) \datac, which consists of large-scale commits suitable for fine-tuning and pretraining, and (2) \datab and \datav, which contain high-quality, human-confirmed bugs and vulnerabilities, suitable for rigorous evaluation of APR techniques.

\subsection{Raw Data Collection and Filtering} \label{sec:plausible-commit}
\textbf{Commit Collection (38 million).} 
To identify buggy functions from real-world C/C++ projects, we follow established practices in prior work~\cite{zhou2021spi,defect4c_ase_demo:BUGSC} and collect bug-related commits from GitHub repositories. We primarily leverage BigQuery to extract relevant commits based on the following criteria:

\begin{itemize}[leftmargin=*]
\item Projects are open-source, non-fork C/C++ repositories with redistributable licenses;
\item Commits are dated between January 2015 and Dec 2023, sourced from the GH Archive~\cite{gharchive:tool};
\item Projects must have at least 200 stars, indicating a minimum threshold of popularity and community engagement.
\end{itemize}
Using these criteria, we identified 110,441 candidate repositories. Due to resource constraints and to ensure relevance, we retained the top 500 C/C++ repositories ranked by GitHub stars~\cite{githubranking}. To isolate bug-related commits, we employed a keyword-based heuristic filtering approach inspired by VRepair~\cite{chen2022neural:tse}. Specifically, we considered a commit as potentially bug-related if its message contained any of the following keywords: \textit{fix}, \textit{solve}, \textit{repair}, \textit{bug}, \textit{issue}, \textit{problem}, \textit{error}, \textit{fault} and \textit{vulnerability}.
Using this method, we extracted over \textbf{38 million} commits across the selected repositories. 

While the 38 million bug-related commits provide a broad foundation, it is non-trivial to determine whether these commits correspond to actual vulnerabilities or general bugs. To specifically incorporate known vulnerabilities into our benchmark, we further curated a vulnerability-focused dataset by collecting Common Vulnerabilities and Exposures (CVEs) related to C/C++ programming from the CVEProject repository\footnote{\url{https://github.com/CVEProject/cvelist}}, which contains records spanning from 1999 to 2024.
We selected only those CVEs that explicitly provided a single patched commit ID, resulting in a total of \textbf{14,488} vulnerability-related commits. This selection criterion was adopted for two key reasons: (1) CVEs with a single commit ID allow precise retrieval of the vulnerable code changes, enabling accurate identification of the affected functions; and (2) CVEs associated with multiple commits introduce ambiguity, making it difficult to determine which specific change addressed the vulnerability.

In total, our raw dataset consists of approximately \textbf{38+ million }commits, comprising 38 million bug-related commits and 14.5K vulnerability-related commits, forming the foundation for further refinement and construction of our benchmark.

\textbf{Commit Validation (9 million).} 
We recognize that some of the commits collected from BigQuery and the CVE repository may become unavailable or invalid over time due to factors such as repository ownership changes, archival, or deletion.  To ensure data integrity, we apply a rigorous filtering and deduplication process. The criteria are as follows: (1) exclude inaccessible or privatized repositories; (2) exclude repositories transferred from highly starred owners to lower-ranked ones or restricted by newly imposed licenses; (3) remove forks with largely duplicated commit histories (e.g., apple/clang forked from llvm/llvm-project, where commits differ only by SHA but not by content); (4) eliminate redundancy by removing MD5-hash duplicates in both source-code patch diffs and corresponding test-case diffs; (5) filter files with non-C/C++ extensions; (6) exclude commit hashes not recognized in GitHub repositories; and (7) discard commits whose buggy–patched diffs are excessively large, as they are more indicative of general function updates or refactoring rather than targeted bug fixes.



This results in a refined dataset of approximately 9 million valid bug-relevant commits.
From these commits, we extract function-level code pairs, specifically, the function before and after the commit, which represent the potential buggy and patched versions, respectively. These examples are particularly valuable for fine-tuning or pretraining APR models, especially given the absence of large-scale real-world C/C++ bug repair datasets for learning-based approaches.
However, these commits are not suitable for rigorous evaluation due to two main limitations: (1) they may include false positives, such as commits unrelated to actual bug fixes (e.g., refactoring or minor edits), and (2) some lack associated unit tests or reproducible setups to verify the correctness of the fix. As such, they serve primarily as training resources, rather than rigorous evaluation benchmarks.

\textbf{Single-Function Commit Filtering (76K).}
The initially collected commits often involve changes across multiple files or functions, which pose challenges for existing APR techniques that typically focus on single-line, single-hunk, or single-function bugs~\cite{xia2023keep,kong2024contrastrepair}. 
Specifically, \textit{Line} refers to bugs where the fix is confined to a single line of code; \textit{Hunk} represents fixes involving multiple consecutive lines (i.e., a continuous code block); and \textit{Function} encompasses fixes that involve non-contiguous changes within a single function.
To reduce complexity and align with the capabilities of current repair models, we retain only those commits that modify exactly one function. Furthermore, to ensure that the extracted functions are executable, which is necessary for validating the correctness of the fix, we filter out commits that lack an associated test suite for validation.

Applying these criteria, we identify a refined set of 76K valid single-function commits, which includes 249 commits linked to known vulnerabilities. This curated subset offers a more controlled and evaluable environment for function-level program repair research.


\subsection{Unit Test Extraction and Matching} \label{sec:pair-verification}
To validate the correctness of fixes, we extract unit tests that can be used to test whether a patch is plausible, i.e., whether it causes the program to pass its test cases. However, after the commit validation and filtering process described in Section~\ref{sec:plausible-commit}, each commit is typically associated with a test suite containing multiple test cases, many of which are designed to validate general functionality rather than the specific bug fix in the commit. Therefore, we need to identify the specific test cases that evaluate the targeted fix.

While simple heuristics exist in other ecosystems, for example, in Java, where a function named \texttt{abc} is often tested by a test named \texttt{test\_abc}, such naming conventions are infrequently used in C/C++ projects, rendering this approach ineffective. To address this, we propose a unit test pair verification algorithm based on a key observation: \textit{for a genuine bug or vulnerability fix, there typically exists at least one unit test that passes on the corrected version but fails on the buggy version.}
Formally, let the test suite be denoted as $T = {(t_1, t_2, ..., t_n)}$, and let a commit produce two versions of code: $V_0$ (pre-commit) and $V_1$ (post-commit). For each test case $t_i \in T$, we execute $t_i$ on both versions. If $t_i$ passes on $V_1$ but fails on $V_0$, we consider it a bug-revealing test case that is directly associated with the fix. We discard test cases that do not show this behavioral difference, as they are unlikely to be related to the fix. The resulting subset $T' \subseteq T$ includes only the test cases that specifically validate the buggy function.

By applying this test verification process to the 76K candidate commits from Section~\ref{sec:plausible-commit}, we identify a high-quality subset consisting of \textbf{3,785} commits for \datab and \textbf{\datasetsizecve} commits for \datav, both of which include executable buggy functions and corresponding bug-revealing test cases.

\subsection{Human Confirmation and Bug Classification} \label{sec:human-annotation}

Given the potential presence of false positives in both the bug-related commits and the associated unit tests, we conducted a conservative and rigorous human annotation process to ensure the construction of a high-quality evaluation dataset for APR techniques. Each commit and its corresponding test cases were manually analyzed by human experts, who reviewed the code and commit messages, executed the unit tests, and thoroughly understood the program logic to determine: (1) whether the change was genuinely bug-related, (2) whether the associated unit test was relevant, and (3) the type of bug in terms of its root cause.

Following the methodology of prior studies~\cite{quan2022towards,shi2022large}, we applied a multi-round annotation protocol to the 3,785 general commits and 102 vulnerability-related commits identified in Section~\ref{sec:pair-verification}. The dataset was first randomly divided into two equal halves, and annotated in successive rounds. In the first round, half of the dataset was independently labeled by two experienced annotators, each with at least 5 years of programming experience and over 3 years in software testing or program analysis. The annotators then discussed their annotations to resolve discrepancies, with final decisions adjudicated by an independent arbitrator.
In the second round, the remaining half of the dataset was annotated using the agreed-upon guidelines. To further ensure reliability, we performed a third round involving a resampling and re-verification of the entire dataset. Only commits confirmed to be genuinely bug-related and paired with valid unit tests were retained.

To evaluate inter-annotator agreement, we used Cohen’s Kappa ($\kappa$) coefficient\cite{hsu2003interrater}, a standard measure of inter-rater reliability. In the first round, the $\kappa$ value was 0.48, indicating moderate agreement. After refining the annotation taxonomy and the criteria, the second round achieved a $\kappa$ of 0.70. Finally, in the third round, after additional consensus-building discussions and verification, the $\kappa$ score improved to 0.88, which is considered almost perfect agreement~\cite{landis1977measurement}. At this point, further rounds of annotation were deemed unnecessary.

During this process, we discovered that some commits, despite containing bug-related keywords, were unrelated to actual bugs, instead introducing new features or modifying output formats. Others had vague messages (e.g., ``fix bug'') that were inconsistent with the code changes, or were later reverted, further calling into question their reliability. After completing the annotation process, we curated \textbf{\datasetsizenorm} high-confidence general bug commits for \datab and retained \textbf{\datasetsizecve} vulnerability-related commits for \datav. Notably, no vulnerability commits were removed, as they originated from the high-quality, curated CVE repository. In total, we obtained \textbf{350} high-quality, reproducible faults, each paired with a corresponding unit test, making them well-suited for rigorous evaluation of APR techniques.

\section{Statistics of \dataset} \label{sec:statistics}
Finally, \dataset comprises a total of 9 million bug-related commits under \datac, including 76,000 single-function commits with potential test suites and 3,887 commits with executable test cases. From this refined set, we identified \datasetsizenorm confirmed general bugs for \datab and \datasetsizecve confirmed vulnerabilities for \datav through rigorous human validation. Note that the 350 confirmed bugs serve as rigorous benchmarks for evaluating APR techniques, similar to Defects4J. These high-quality, reproducible faults, each paired with executable test cases, are suitable for use in empirical studies and comparative evaluations. In addition to this evaluation subset, the remaining commits, with function-level before-and-after pairs, offer a valuable resource for fine-tuning or pretraining APR models. Users may further apply customized filtering or preprocessing to tailor the data to their specific fine-tuning objectives, such as selecting by project domain, filtering by commit metadata, or augmenting with different strategies. 

Table~\ref{dataset-types} presents the taxonomy and statistical summary of the confirmed bugs, categorized based on their error types as determined through manual analysis. 
The dataset is classified into four primary categories based on the logical location of the fix: \texttt{Signature}, \texttt{Sanitizer}, \texttt{Memory Error}, and \texttt{Logic Organization}. Each primary category is further divided into subcategories, reflecting more fine-grained root causes and bug patterns observed during annotation.
Due to space limitations, we provide a detailed description of the full taxonomy and examples for each category on our project website~\cite{ourweb}. 

Furthermore, we classify the bug-fix patterns in \dataset based on the granularity of code modifications, dividing them into three categories: \textit{Line}, \textit{Hunk}, and \textit{Function}.  This categorization provides insights into the structural complexity of the fixes and helps guide the design of APR models with appropriate capabilities. A detailed breakdown of the error distribution across these three categories for various C/C++ projects is available on our project website~\cite{ourweb}.

\textbf{Usage.}
We recognize that usability is a critical requirement for datasets supporting research in areas such as program repair and vulnerability detection (e.g., Defects4J). To maximize usability for the research community, we developed a stateless HTTP and command-line interface (CLI) designed to support large-scale automated program repair evaluation. This interface addresses three key challenges: (1) scalable end-to-end patch extraction and verification, (2) isolated verification environments, and (3) efficient integration with large language models, including compatibility with their generated responses.

The interface exposes two primary endpoints. The first, \verb|/extract_anchor_patch|, extracts patches from raw LLM outputs, identifies corresponding anchor points, and integrates the patches into the source code. while \verb|/fix_with_patch| performs isolated patch verification by applying patches within a Docker container (all bugs co-exist in one container) and returning a Boolean success status along with categorized error feedback for failed attempts. To support high-throughput use, we implement dual caching strategies, a Redis web cache and a C/C++ builder cache, to efficiently manage millions of concurrent and repeated requests.

We also provide additional tools to enhance the debugging and verification experience. The \verb|/reproduce| endpoint resets and reinitializes the verification environment for a given bug, while the \verb|/error_dig| interface performs structured error analysis by classifying failures (e.g., compile, build, link, or test), identifying root causes, and locating error positions via stack trace parsing. The output is formatted to be LLM-friendly, particularly under context-length constraints. Detailed implementation guidance and usage documentation are available on our project website~\cite{ourweb}.

\begin{table}[]

\centering
\caption{The number of bugs and vulnerabilities across categories.~\label{dataset-types}}
\resizebox{0.5\textwidth}{!}{
\setlength{\tabcolsep}{1mm}{
\begin{tabular}{l|c|cc}
\toprule 
\multicolumn{1}{l|}{Category} & \multicolumn{1}{c|}{Error Type} & \multicolumn{1}{c}{Bugs} & \multicolumn{1}{c}{Vulnerabilities}\tabularnewline
\toprule
\multirow{4}{*}{Signature } & Incorrect Function Usage & 19 & 3 \tabularnewline
 &  Fault Input Type & 12 & 2 \tabularnewline
 &  Incorrect Function Return Value & 19 & 3 \tabularnewline
 &  Incorrect Variable Usage & 25 & 3\tabularnewline
\midrule
Sanitizer  & Control Expression Error & 66 & 6 \tabularnewline
\midrule
\multirow{3}{*}{Memory Error } & Null Pointer Dereference & 6 & 6 \tabularnewline
 &  Uncontrolled Resource Consumption & 9 & 5\tabularnewline
 &  Memory Overflow & 5 & 61 \tabularnewline
\midrule
\multirow{2}{*}{Logic Organization } & Improper Condition Organization & 67 & 11 \tabularnewline
 & Wrong Function Call Sequence & 20 & 2 \tabularnewline
 \bottomrule
\end{tabular}
}
\vspace{-8mm}
}
\end{table}

\section{Evaluation}


\subsection{Evaluation Workflow}

Large language models~(LLMs) have demonstrated significant potential in APR~\cite{xia2023automated,xia2023conversational,xia2023keep} with competitive or even better performance compared to traditional techniques. However, existing works mainly focus on evaluating the APR effectiveness of LLMs on Java and Python projects, neglecting their repair capability on C/C++. To bridge this gap, in this work, we conduct a comprehensive empirical study to evaluate the performance of LLMs on C/C++ program repair tasks using our constructed \dataset dataset.

In particular, our study contains two parts. First, we directly employ pre-trained LLMs to fix bugs hidden in our evaluation datasets \datab and \datav to assess their program repair ability. Here, we consider different LLM-based program repair strategies: i.e., single-round repair and conversation-based repair. 

\textit{Single-round repair} refers to the model generating a patched program once based on the given prompt, without receiving feedback or undergoing multiple iterations of verification and re-generation, which is a basic strategy for LLM-based APR. 
\textit{Conversation-driven repair}, as proposed by Xia et al.~\cite{xia2023keep}, involves iteratively invoking the model multiple times. In each iteration, the generated program will be executed by a provided compiler and corresponding test cases. If the program is not executable or pass, the error feedback will be incorporated into the prompt and used in the next iteration as guidance to help generate correct programs. This strategy contains two hyperparameters, $m$ and $n$, representing the maximum number of repair attempts and the maximum conversation length in each attempt.


Second, the majority of LLM-based APR research relies on pre-trained models, primarily due to the lack of datasets capable of supporting large-scale fine-tuning for repair tasks. However, our dataset \datac addresses this limitation. Therefore, we further conduct a study to evaluate the repair performance of LLMs with fine-tuning. Specifically, we select single-function commits paired with test suites from \datac as the fine-tuning dataset and evaluate the performance of the fine-tuned models on \datab and \datav. Following the approach used in Magicoder~\cite{wei2023magicoder}, we perform decontamination to exclude any samples that are identical to, or share similar buggy or patched code snippets with, those in \datab and \datav to prevent data leakage. This was achieved by employing UniXcoder~\cite{guo2022unixcoder} to embed code snippets and filtering out samples with a cosine similarity score higher than 0.95 when compared to samples in \datab and \datav. After filtering the input length greater than 2048, we retained 20,591 samples from \datac across 1.1K projects for fine-tuning. By comparing the results before and after fine-tuning, we investigate the usefulness of our dataset to boost the program repair capability of LLMs. 

We plan to answer the following research questions in the study:

\textbf{RQ1: } \textit{How effective are pre-trained LLMs in fixing bugs in \dataset?}

\textbf{RQ2: } \textit{How does LLMs perform on APR tasks after fine-tuning with \dataset?}

\textbf{RQ3: } \textit{What are the characteristics of errors made by LLMs on \datav?}




\subsection{Prompt Design}

\begin{figure*}[ht]
\centering
     \includegraphics[width=0.93\textwidth]{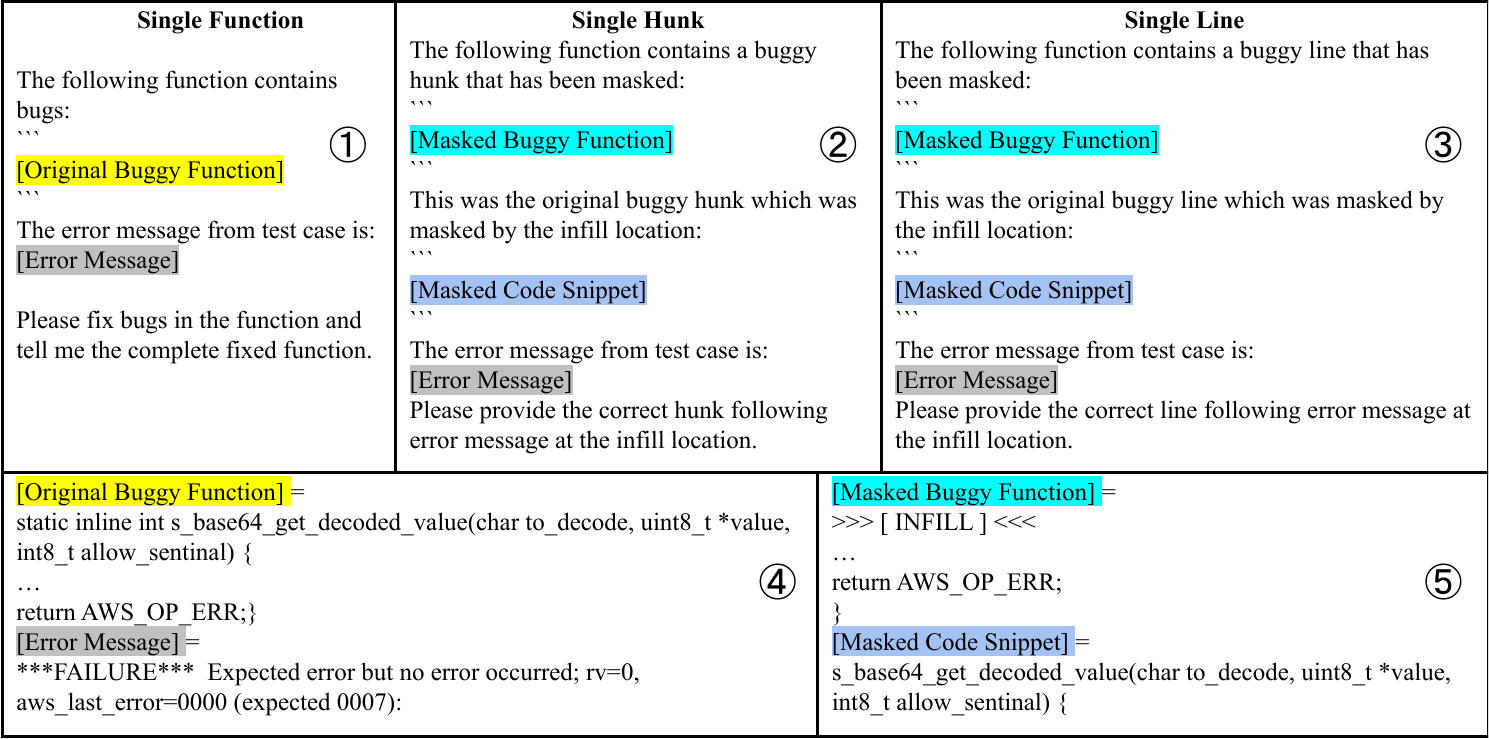}
     \caption{Prompt design for different types of defects.}
     \label{fig:prompt_design}
\end{figure*}

To interact with LLMs, we need to design appropriate input prompts. Based on the three types of bugs/vulnerabilities, i.e., fixed in a single line, hunk, or function, as described in Section~\ref{sec:statistics}, we design corresponding prompts respectively. Figure~\ref{fig:prompt_design} illustrates the prompt templates. For single function bugs, we design prompts to require the model to generate the complete function. A concrete example is given in the part \ding{175} of Figure~\ref{fig:prompt_design}. For the prompt for the single hunk and single line bugs, as the error statements are continuous, we mask them in the original function by the symbol $>>>$[INFILL]$<<<$ and provide these error statements by the placeholder \texttt{Masked Code Snippet} for the model to generate masked statements. An example is shown in the part \ding{176} of Figure~\ref{fig:prompt_design}.


For single-round repair, we directly feed the prompts to the model. For conversation-based repair, the designed prompts serve as the initial input to the LLMs. After the model generates an output, the compiler evaluates it. If the output fails to pass the verification, the newly produced error feedback is appended to the prompt template to construct a new prompt for the next round of repair. 
For fine-tuning, we use the same prompt as the single-round repair for the evaluation.

\subsection{Experimental Setup}

\textbf{Subject LLMs.} For RQ1, our evaluation considers \llmsize types of pre-trained LLMs, covering almost all famous LLMs such as GPT-4, CodeLlama, and DeepSeek. The detailed LLMs used can be found in Table~\ref{tbl-program-overall}. For RQ2, due to resource constraints, we select two popular open-source models, CodeLlama-7B-base and DeepSeek-coder-6.7B-base, for fine-tuning.

\textbf{Evaluation Metrics.} For the single-round repair evaluation, we follow EvalPlus~\cite{evalplus} and use unbiased pass@$k$~\cite{chen2021codex:humaneval} to assess the repair capacity of LLM. Here, we set $k$ as 1, 10, and 100. For conversation-based repair, it is costly to use pass@$k$ in this setting, since pass@$k$ requires generating a massive amount of model outputs. Hence, we follow Xia et al.~\cite{xia2023keep} to report the number of successful repairs in \dataset. 

\textbf{Configuration.} For single-round repair, we set model temperature as 0.2 and 0.8. For greedy-search decoding, we follow~\cite{evalplus} to evaluate its pass rate as pass@$k^*=1$. GPT-4 is only evaluated under greedy decoding due to time and cost constraints. For conversation-based repair, we follow~\cite{xia2023keep} to set model temperature as 1.0. In our conversation-based repair experiments, we compare two decoding strategies distinguished by determinism: deterministic~(greedy) greedy-search decoding~($T=0$); and non-deterministic decoding~($T=1$), which samples from the full probability distribution, introducing stochasticity and enhancing output diversity. Our default configuration uses up to 10 repair attempts with a conversation length limited to 3 turns per attempt, resulting in a total budget of 30 repair steps per buggy function; the process terminates when an output passes all test cases or the 30-step budget is exhausted. For LLM fine-tuning, we apply parameter-efficient fine-tuning using LoRA~\cite{hu2021lora} with a rank of 8. The models are fine-tuned for 3 epochs with a learning rate of 2e-5. The batch size is 16, and the maximum input sequence length is 2048 for all experiments.

\textbf{Environments.} All experiments are conducted on a server with 8X A100-SXM4-80GB GPUs. More detailed settings on our project website~\cite{ourweb}.

\section{Experimental Results}


\begin{table*}[!t]
\caption{Evaluating LLMs on \dataset for conversation-based repair where Pass denotes the number of bugs or vulnerabilities that the model can successfully repair, Avg.tries denotes the average tries of the successful repair. Due to the limited budget, the maximum number of repair attempts is set to 2 for GPT-4, and the remaining models are set to 10 by default.}
\label{dataset-conversation}
\setlength{\tabcolsep}{1mm}{
\resizebox{1\textwidth}{!}{
\centering

\begin{tabular}{ll|cccccccc|c||cccccccc|c}
\toprule 
\multirow{3}{*}{Model} & \multirow{3}{*}{Decoding} & \multicolumn{8}{c|}{\datab} & \multirow{3}{*}{Pass/Sum} & \multicolumn{8}{c|}{\datav} & \multirow{3}{*}{Pass/Sum}\tabularnewline
 &  & \multicolumn{2}{c}{Signature} & \multicolumn{2}{c}{Sanitizer} & \multicolumn{2}{c}{Memory Error} & \multicolumn{2}{c|}{Logic} & & \multicolumn{2}{c}{Signature} & \multicolumn{2}{c}{Sanitizer} & \multicolumn{2}{c}{Memory Error} & \multicolumn{2}{c|}{Logic} & \tabularnewline
 &  & Pass/Total & Avg.tries & Pass/Total & Avg.tries & Pass/Total & Avg.tries & Pass/Total & Avg.tries & & Pass/Total & Avg.tries & Pass/Total & Avg.tries & Pass/Total & Avg.tries & Pass/Total & Avg.tries & \tabularnewline
\midrule 
\multirow{2}{*}{GPT-4} 
 & T=1.0 & 0/75 & 0.0 & 4/66 & 2.0 & 1/20 & 1.0 & 0/87 & 0.0 & 5/\datasetsizenormi & \textbf{1}/11 & 2.0 & 0/6 & 0 & 4/72 & 1.5 & 0/13 & 0.0 & 5/\datasetsizecvei\tabularnewline
 & greedy & 3/75 & 2.0 & 1/66 & 1.0 & 1/20 & 2.0 & 0/87 & 0.0 & 5/\datasetsizenormi & \textbf{1}/11 & 2.0 & 0/6 & 0.0 & 3/72 & 1.3 & 0/13 & 0.0 & 4/\datasetsizecvei\tabularnewline
\midrule 
\multirow{2}{*}{GPT-35-Turbo} 
 & T=1.0 & 8/75 & 1.7 & \textbf{13}/66 & 2.4 & 3/20 & 3.7 & 3/87 & 2.7 & \textbf{27}/\datasetsizenormi & 0/11 & 0.0 & 1/6 & 10.0 & 0/72 & 0.0 & 0/13 & 0.0 & 1/\datasetsizecvei\tabularnewline
 & greedy & 7/75 & 2.0 & 4/66 & 3.0 & \textbf{5}/20 & 2.8 & 2/87 & 1.0 & 18/\datasetsizenormi & 0/11 & 0.0 & \textbf{2}/6 & 4.5 & 2/72 & 8.5 & 0/13 & 0.0 & 4/\datasetsizecvei\tabularnewline
\midrule 
\multirow{2}{*}{CodeLlama-Instruct-7B} 
 & T=1.0 & \textbf{9}/75 & 2.8 & 11/66 & 2.9 & 3/20 & 3.0 & 4/87 & 6.3 & \textbf{27}/\datasetsizenormi & 0/11 & 0.0 & 0/6 & 0.0 & 0/72 & 0.0 & 0/13 & 0.0 & 0/\datasetsizecvei\tabularnewline
 & greedy & 3/75 & 6.0 & 8/66 & 4.6 & 4/20 & 4.7 & 1/87 & 1.0 & 16/\datasetsizenormi & 0/11 & 0.0 & 0/6 & 0.0 & 0/72 & 0.0 & \textbf{1}/13 & 9.0 & 1/\datasetsizecvei\tabularnewline
\midrule 
\multirow{2}{*}{Gemma-Instruct-7B} 
 & T=1.0 & 0/75 & 0.0 & 1/66 & 1.0 & 0/20 & 0.0 & 0/87 & 0.0 & 1/\datasetsizenormi & 0/11 & 0.0 & 0/6 & 0.0 & 1/72 & 3.0 & 0/13 & 0.0 & 1/\datasetsizecvei\tabularnewline
 & greedy & 1/75 & 8.0 & 0/66 & 0.0 & 0/20 & 0.0 & 0/87 & 0.0 & 1/\datasetsizenormi & 0/11 & 0.0 & 0/6 & 0.0 & 0/72 & 0.0 & 0/13 & 0.0 & 0/\datasetsizecvei\tabularnewline
\midrule 
\multirow{2}{*}{WizardCoder-Python-34B} 
 & T=1.0 & 0/75 & 0.0 & 0/66 & 0.0 & 0/20 & 0.0 & 1/87 & 1.0 & 1/\datasetsizenormi & 0/11 & 0.0 & 0/6 & 0.0 & 0/72 & 0.0 & 0/13 & 0.0 & 0/\datasetsizecvei\tabularnewline
 & greedy & 0/75 & 0.0 & 0/66 & 0.0 & 0/20 & 0.0 & 0/87 & 0.0 & 0/\datasetsizenormi & 1/11 & 8.0 & 0/6 & 0.0 & 0/72 & 0.0 & 0/13 & 0.0 & 1/\datasetsizecvei\tabularnewline
\midrule 
\multirow{2}{*}{Phind-CodeLlama-34B} 
 & T=1.0 & \textbf{9}/75 & 4.9 & 4/66 & 6.7 & 1/20 & 8.0 & 4/87 & 4.7 & 18/\datasetsizenormi & 0/11 & 0.0 & \textbf{2}/6 & 1.0 & \textbf{5}/72 & 4.8 & 0/13 & 0.0 & \textbf{7}/\datasetsizecvei\tabularnewline
 & greedy & 0/75 & 0.0 & 2/66 & 1.0 & 4/20 & 1.0 & 1/87 & 8.0 & 7/\datasetsizenormi & 0/11 & 0.0 & 1/6 & 1.0 & 1/72 & 1.0 & 0/13 & 0.0 & 2/\datasetsizecvei\tabularnewline
\midrule 
\multirow{2}{*}{deepseek-coder-33b-base} 
 & T=1.0 & 4/75 & 1.5 & 0/66 & 0.0 & 2/20 & 1.0 & 0/87 & 0.0 & 6/\datasetsizenormi & 0/11 & 0.0 & 0/6 & 0.0 & 0/72 & 0.0 & 0/13 & 0.0 & 0/\datasetsizecvei\tabularnewline
 & greedy & 0/75 & 0.0 & 0/66 & 0.0 & 0/20 & 0.0 & \textbf{6}/87 & 8.2 & 6/\datasetsizenormi & 0/11 & 0.0 & 0/6 & 0.0 & 0/72 & 0.0 & 0/13 & 0.0 & 0/\datasetsizecvei\tabularnewline
\bottomrule
\end{tabular}
\vspace{-4mm}
}
}
\end{table*}

\subsection{RQ1: Effectiveness of Pre-Trained LLMs on \dataset}



\textbf{Single-round repair evaluation.} The single-round repair results of different LLMs on \dataset are presented in Table~\ref{tbl-program-overall}. First, we can conclude that LLMs with temperature 0.8 usually outperform LLMs with temperature 0.2 in this APR task. This indicates that increasing the diversity of model outputs leads to better program repair capability of LLMs. Further analysis of different variants of the same model reveals that increasing model size does not necessarily lead to better repair accuracy. For instance, when the size of CodeLlama-Python increases from 7B to 13B, pass@100 improves from 22.5 to 32.2. However, with CodeLlama-Python 34B, pass@100 drops to 29.8. Similar trends are observed in WizardCoder-15B/33B and CodeLlama-Instruct. {We conducted an in-depth analysis to understand this surprising results and found that larger models tend to generate more verbose and detailed outputs, including lengthy explanations before or alongside the patch. While these additional explanations may reflect stronger reasoning ability, they also lead to practical issues: In some cases, the verbose output exceeds the token limit (2048 tokens, following EvalPlus), resulting in incomplete patches, approximately 19\% of cases for CodeLlama-Instruct-34B failed to produce complete patched output due to such overgeneration. Furthermore, the over-explanation increases the likelihood of hallucinations in some cases, which could inadvertently degrade the correctness of the generated code. } Besides, the performance gap between open-source and closed-source models on \dataset is less pronounced compared to their performance on other datasets~\cite{chen2021codex:humaneval}. This indicates that \dataset, collected from real-world projects, presents a more challenging testbed.


\begin{table*}[!t]
\centering
  \renewcommand{\arraystretch}{1.05} 
  \renewcommand{\tabcolsep}{5pt} 
\caption{Evaluating LLMs on \dataset for single-round repair, where $k^*=1$ marks pass@1 done with greedy-search decoding and pass@$k$ results with its corresponding temperature.}
\vspace{-0.5mm}
  \resizebox{0.65\textwidth}{!}{%

\small
\label{tbl-program-overall}

\begin{tabular}{ll|r|rrr|ccc}
\toprule
\multirow{2}{*}{Model} & \multirow{2}{*}{Size}&  & \multicolumn{3}{c|}{T=0.2} & \multicolumn{3}{c}{T=0.8}\tabularnewline
& & k{*}=1 &$k=1$ & $k=10$ & $k=100$ &  $k=1$ & $k=10$ & $k=100$ \tabularnewline
\midrule

GPT-4 & N/A & \textbf{9.0} & - & - & - & - & -  & - \tabularnewline
GPT-35-Turbo & N/A  & 8.5 & 7.9 & 13.5 & 19.5 & 7.1 & 20.0 & 38.9\tabularnewline
\midrule 
\multirow{3}{*}{CodeLlama-Python} & 7B & 0.0 & 0.1 & 1.2 & 4.5 & 0.8 & 6.2 & 22.5\tabularnewline
 & 13B & 0.0 & 0.3 & 1.8 & 4.5 & 1.7 & 11.2 & 32.2\tabularnewline
 & 34B & 0.0 & 0.3 & 2.2 & 6.9 & 1.2 & 8.8 & 29.8\tabularnewline
\midrule 
CodeLlama-Base  & 7B & 0.0 & 0.0 & 0.0 & 0.0 & 0.2 & 2.1 & 14.3\tabularnewline

\midrule 
\multirow{3}{*}{CodeLlama-Instruct} & 7B & 2.5 & 3.3 & 11.1 & 24.9 & 4.8 & 20.5 & 45.7\tabularnewline
 & 13B & 5.3 & 4.0 & 14.2 & 25.7 & 3.8 & 18.1 & 40.4\tabularnewline
 & 34B & 4.0 & 3.6 & 12.1 & 25.7 & 3.2 & 14.7 & 35.9\tabularnewline

\midrule
\multirow{3}{*}{deepseek-coder} & 6.7B-Inst. & 1.2 & 2.4 & 10.7 & 25.7 & 2.2 & 13.4 & 33.9\tabularnewline
& 6.7B & 0.6 &  0.0 & 0.0 & 0.0 & 0.5 & 3.8 & 12.2\tabularnewline
&  33B &0.3 & 0.4 & 1.1 & 2.4 & 1.4 & 8.7 & 21.6\tabularnewline

\midrule 
\multirow{3}{*}{Gemma}  &7B-Inst.  & 0.0  & 0.8 & 5.1 & 14.7 & 0.9 & 6.1 & 22.9\tabularnewline
& 7B & 0.0 & 0.4 & 3.0 & 11.0 & 0.8 & 6.6 & 26.9\tabularnewline
 &Code7B   & 0.0 & 0.0 & 0.0 & 0.0 & 0.0 & 0.2 & 1.2\tabularnewline
\midrule 
phi-2 & 2.7B & 0.0 & 0.0 & 0.0 & 0.0 & 0.4 & 3.7 & 19.9 \tabularnewline
Magicoder-S-DS & 6.7B & 3.3 & 2.6 & 9.9 & 24.7 & 4.7 & 22.6 & 34.8 \tabularnewline
Mixtral-8x7B-Instruct & 7B & 0.0 & 1.5 & 7.2 & 16.2 & 1.6 & 7.4 & 13.1\tabularnewline
\midrule 
Phind-CodeLlama & 34B & 6.1 & 5.4 & 18.6 & 34.7 & 4.8 & 20.6 & 38.4\tabularnewline
\midrule 
\multirow{3}{*}{WizardCoder-Python} & 7B & 0.0  & 0.2 & 1.1 & 3.7 & 0.4 & 3.4 & 18.8\tabularnewline
 & 13B & 0.0  & 0.7 & 4.2 & 11.8 & 1.4 & 11.0 & 35.5\tabularnewline
 & 34B & 4.4  & 5.2 & 13.0 & 21.2 & 5.5 & 23.0 & 45.1\tabularnewline
\midrule 
\multirow{2}{*}{WizardCoder} & 15B & 1.0 & 1.1& 4.9 & 11.3 & 1.7 & 10.4 & 28.9\tabularnewline
 & 33B & 0.0 & 5.5 & 6.8 & 11.0 & 3.2 & 10.7 & 18.9\tabularnewline

\bottomrule
\end{tabular}

} 
\end{table*}

\textbf{Conversation-based repair evaluation.}  We then select the best performing models from Table~\ref{tbl-program-overall} to perform experiments on conversation-based repair, with the results presented in Table~\ref{dataset-conversation}. The first conclusion we can draw is that LLMs perform better in repairing \datab than \datav. The best LLMs can successfully repair 27 bugs in \datab, while only 7 vulnerabilities in \datav. We conjecture that this difference comes from the increased complexity of vulnerabilities,  making them more difficult for LLMs to address. However, considering the total number of bugs~(248) and vulnerabilities~(102), the success repair rate for \datab and \datav are only 10.88\% and 6.86\%, respectively. This low performance highlights the significant room for improvement in LLMs' ability to repair C/C++ defects.  


Additionally, since we limit the repair attempts of GPT-4 to 2 due to the budget constraints, it performs worse than GPT-3.5 on \datab. However, GPT-4 demonstrates potential on \datav, with the second-best repairing performance. We believe that GPT-4 could achieve higher repair accuracy with more repair attempts. Lastly, apart from GPT-4 and GPT-3.5, open-source models perform poorly even in conversation-based repair. For example, WizardCoder and Gemma are able to repair only 1 bug or vulnerability in both \datab and \datav. This suggests that while these open-source models may excel in certain tasks or datasets reported by existing works, their generalizability remains limited.


\begin{table}[!t]
\centering
\caption{The repair performance compared with Defects4J. \#Avg.tries represents the average number of attempts required, calculated as the ratio of successful repairs (Pass) to the total attempts (Total).}
\vspace{2mm}
\resizebox{0.99\linewidth}{!}{
\setlength{\tabcolsep}{1mm}{
\small
\label{compared-defect4j}

\begin{tabular}{lc|cc|cc|cc|c}
\toprule
\multicolumn{2}{c|}{Model} & \multicolumn{2}{c|}{Func} & \multicolumn{2}{c|}{Hunk} & \multicolumn{2}{c|}{Line} & \multirow{2}{*}{\#Avg.tries}\tabularnewline
\multicolumn{2}{c|}{} & \#Pass/Total & Rate & \#Pass/Total & Rate & \#Pass/Total & Rate & \tabularnewline
\midrule 
\multicolumn{2}{c|}{Defects4J~\cite{xia2023keep}} & - & 29.80 & - & 51.30 & - & 71.30 & - \tabularnewline
\midrule 
\multirow{2}{*}{GPT4} & T=1 & 1/46 & 2.17 & 2/179 & 1.12 & 7/125 & 5.60 & 2.86\tabularnewline
 & greedy & 0/46 & 0.00 & 7/179 & 3.91 & 2/125 & 1.60 & 2.57\tabularnewline
\midrule 
\multirow{2}{*}{GPT-3.5-Turbo} & T=1 & 0/46 & 0.00 & 11/179 & 6.15 & 17/125 & 13.60 & 8.00\tabularnewline
 & greedy & 0/46 & 0.00 & 9/179 & 5.03 & 13/125 & 10.40 & 6.29\tabularnewline
\midrule 
\multirow{2}{*}{CodeLlama-Instruct-7B} & T=1 & 0/46 & 0.00 & 10/179 & 5.59 & 17/125 & 13.60 & 7.71\tabularnewline
 & greedy & 0/46 & 0.00 & 10/179 & 5.59 & 7/125 & 5.60 & 4.86\tabularnewline
\midrule 
\multirow{2}{*}{Gemma-Instruct-7B} & T=1 & 0/46 & 0.00 & 1/179 & 0.56 & 1/125 & 0.80 & 0.57\tabularnewline
 & greedy & 0/46 & 0.00 & 1/179 & 0.56 & 0/125 & 0.00 & 0.29\tabularnewline
\midrule 
\multirow{2}{*}{WizardCoder-Python-34B} & T=1 & 0/46 & 0.00 & 1/179 & 0.56 & 0/125 & 0.00 & 0.29\tabularnewline
 & greedy & 0/46 & 0.00 & 1/179 & 0.56 & 0/125 & 0.00 & 0.29\tabularnewline
\midrule 
\multirow{2}{*}{Phind-CodeLlama-34B} & T=1 & 2/46 & 4.35 & 12/179 & 6.70 & 11/125 & 8.80 & 7.14\tabularnewline
 & greedy & 0/46 & 0.00 & 3/179 & 1.68 & 6/125 & 4.80 & 2.57\tabularnewline
\midrule 
\multirow{2}{*}{deepseek-coder-33b-base} & T=1 & 0/46 & 0.00 & 4/179 & 2.23 & 2/125 & 1.60 & 1.71\tabularnewline
 & greedy & 0/46 & 0.00 & 6/179 & 3.35 & 0/125 & 0.00 & 1.71\tabularnewline
\bottomrule
\end{tabular}

}
}
\end{table}

\textbf{Results comparison between Defects4C and Defects4J.} We further compare the difficulty between Defects4C and Defects4J using the repair results of LLM-based methods. The results of Defects4J and \dataset are presented in Table~\ref{compared-defect4j},  where the first row presents the results for Defects4J.
Note that, we directly report the conversation-driven repair results of Defects4J provided by~\cite{xia2023keep} in the table. In their original setting, GPT-3.5 is used as the base model for conversation-driven repair.
Compared with the repair success rate on Defects4J, the success rate in repairing C/C++~(\dataset) bugs and vulnerabilities is significantly lower, underscoring the challenges in fixing C/C++ faults and the need for more advanced and specific repair methods. 

\begin{tcolorbox}[size=title,opacityfill=0.1,breakable]
\noindent
\textbf{Answer to RQ1:} LLM-based APR techniques can only fix 10.88\% and 6.86\% bugs in \datab and \datav, respectively, which are significantly lower than the bug fixing rate on Defects4J, showcasing the challenges of \dataset.   
\end{tcolorbox}

\vspace{-4mm}

\subsection{RQ2: Effectiveness of Fine-Tuned LLMs on \dataset}\label{finetune}

\begin{table}[!t]
\centering
\caption{Comparative Results With/Without  Fine-Tuning.}
\resizebox{0.99\linewidth}{!}{
\setlength{\tabcolsep}{1mm}{
\small
\label{tbl-finetue-appendix}
\begin{tabular}{lc|c|ccc|ccc}
\toprule 
\multirow{2}{*}{Model} & \multirow{2}{*}{Finetune}  & \multirow{2}{*}{Greedy} & \multicolumn{3}{c|}{T=0.2} & \multicolumn{3}{c}{T=0.8}\tabularnewline
& &   &$k=1$ & $k=10$ & $k=100$ &  $k=1$ & $k=10$ & $k=100$ \tabularnewline
\midrule

\multirow{2}{*}{CodeLlama-7B-Base} & \xmark & 0.00 & 0.00 & 0.00 & 0.00 & 0.22 & 2.10 & 14.29\tabularnewline
 & \checkmark & 0.41 & 0.25 & 0.92 & 2.86 & 0.44 & 3.72 & 20.41\tabularnewline
\multirow{2}{*}{CodeLlama-7B-Instruct} & \xmark & 2.45 & 3.31 & 11.07 & 24.90 & 4.81 & 20.51 & 45.71\tabularnewline
 & \checkmark & 4.08 & 4.26 & 9.30 & 17.14 & 4.92 & 20.99 & 46.94\tabularnewline

\midrule 
\multirow{2}{*}{Deepseek-Coder-6.7B-Base} & \xmark & 0.61 & 0.00 & 0.00 & 0.00 & 0.50 & 3.80 & 12.20\tabularnewline
& \checkmark & 3.35 & 1.83& 2.41 & 2.44 & 1.32 & 3.44 & 6.40 \tabularnewline
\multirow{2}{*}{Deepseek-Coder-6.7B-Instruct} & \xmark & 1.22 & 2.42 & 10.65 & 25.71 & 2.16 & 13.36 & 33.88\tabularnewline
 & \checkmark & 3.27 & 3.74 & 10.49 & 20.82 & 3.87 & 18.41 & 41.22\tabularnewline

\bottomrule
\end{tabular}

}
}
\end{table}

We fine-tune open-sourced LLMs using \datac and evaluate the performance of the fine-tuned models on \datab and \datav. The results are presented in Table~\ref{tbl-finetue-appendix}. The second column, \textit{Finetune}, indicates whether the model has been fine-tuned with \datac, where \xmark~represents the results of the pre-trained model (listed here for comparison purposes) and \checkmark~represents the results with LoRA-based fine-tuning. 

Overall, we observe that fine-tuning is a promising way to boost the repair performance of LLMs on C/C++ bugs. In 21 out of 28 cases,  fine-tuned LLMs have higher Pass@k scores than pre-traiend LLMs, with an average relative improvement of 84.89\%. 
However, even with fine-tuning, our studied LLMs still do not perform well in repairing C/C++ bugs. The best model achieved a 4.92 pass@1 score~(CodeLlama-7B-Instruct), which is far from ideal performance. This indicates that more advanced fine-tuning methods to further improve C/C++ program repair are needed.





\begin{tcolorbox}[size=title,opacityfill=0.1,breakable]
\noindent
\textbf{Answer to RQ2:} Fine-tuning with \dataset benefits the repair capability of LLMs on C/C++ bugs, but the improvements are limited. Proposing new, specific fine-tuning methods for \dataset is in need.  
\end{tcolorbox}

\subsection{RQ3: Error Characteristics on \datav}

We found that LLMs perform poorly on \datav. Therefore, we further investigate the underlying bottlenecks that limit their ability to repair bugs in \datav. We first observe that successful repairs often share common patterns: 1) small correct patches confined to 1 to 2 lines, and simple modifications~(e.g., variable renames or type adjustments); 2) buggy code includes multiple test cases, which offer richer feedback and guide the model toward the correct fix.

Furthermore, we analyze the cases in \datav in which LLMs cannot handle correctly and categorize them according to error patterns. After careful manual checking, we summarize four failure patterns that make LLMs difficult to produce correct patches: long/multi-hunk patches, deletion-centric fixes, missing external context, and insufficient test feedback. 

\begin{itemize}
    \item \textit{Long/multi-hunk patches} indicates that the correct patches are long and span multiple functions or lines, but LLMs cannot generate such complex patches.  
    \item \textit{Deletion-centric fixes} refers to correct patches that require removing part of the code snippets, but LLMs rarely perform code removal.
    \item \textit{Missing external context} refers to correct patches that need additional context~(e.g., data structures or global variables) outside the buggy function, but LLMs are unaware of outside information. 
    \item \textit{Insufficient test feedback} indicates that buggy code only provides a single test case, leading to insufficient feedback.
\end{itemize}
Table~\ref{tab:vul-patterns} summarizes the distribution of failure causes in \datav for CodeLlama-7B-Instruct (results for other models are available on our website). The column \textit{Vanilla\%} reports the distribution of failures using the vanilla model without fine-tuning, while \textit{Tuned$\Delta$} indicates the corresponding changes of these failures after fine-tuning.
We can see that most faults happen to \textit{long/multi-hunk patches} and \textit{insufficient test feedback}, which indicates that current LLMs have difficulty handling complex program repair tasks in our dataset. 

We further observe that the effect of fine-tuning is limited, it primarily improves repairs for the deletion-centric fixes and missing external context types, but fails to address other types of failures. This observation provides useful guidance for improving fine-tuning methods, particularly those that better handle long/multi-hunk patches and insufficient test feedback patterns.

\begin{table}[!t]
  \centering
  \caption{Failure Patterns in \datav and the Effects of Fine-Tuning}
  \label{tab:vul-patterns}
  \begin{tabular}{@{}lrr@{}}
    \toprule
    \textbf{Failure Pattern}                & \textbf{Vanilla\%}  & \textbf{Tuned$\Delta$} \\ 
    \midrule
    Long/multi-hunk patches                         & 52.0      & -0.0        \\
    Deletion-centric fixes                           &  9.8        & -2.9        \\
    Missing external context                          & 28.4        & -1.9        \\
    Insufficient test feedback                    &  9.8          & -0.0      \\

    \bottomrule
  \end{tabular}
\end{table}




\begin{tcolorbox}[size=title,opacityfill=0.1,breakable]
\noindent
\textbf{Answer to RQ3:} In \dataset, multi-line bugs and those requiring external information for repair constitute the largest proportion, and they are particularly challenging for LLMs to fix. 
\end{tcolorbox}


\section{Threat to Validity}

While our dataset is significantly more comprehensive than existing C/C++ benchmarks, potential threats to the validity of results remain due to limitations in bug and project collection. To mitigate this, we have made extensive efforts to gather a large volume of data, over 38+ million bug-relevant commits, from a diverse set of real-world, representative projects, within our resource constraints. We applied rigorous and conservative filtering procedures to ensure a reasonable balance between quantity and quality. 

Another limitation arises from our focus on single-function commits. While this design ensures reliable annotation quality, it excludes multi-function or cross-file defects, such as those involving both a function implementation and its declaration. Although this choice simplifies validation and ensures a large set of high-quality defects, it reduces coverage of certain bug categories. We plan to extend the dataset to include multi-function and cross-file bugs in future releases.


Temporal and contamination biases also pose potential threats. Given the popularity of many selected projects, there is a possibility that similar code patterns may appear in the pre-training corpora of LLMs, which could inadvertently inflate performance. However, our results show that LLMs underperform significantly on our dataset, suggesting that memorization and contamination effects could be minimal and would not affect our main conclusions.

Manual annotation may introduce subjective bias. To address this, we employed two independent annotators and measured inter-annotator agreement using Cohen’s Kappa to ensure annotation consistency.

Lastly, the quality of training data used in RQ2 could also affect results. Some data pairs may not be strictly bug-related, which may impact fine-tuning effectiveness. Due to scalability constraints, we did not manually verify each pair. We leave the exploration of improved preprocessing and fine-tuning techniques as future work.

\section{Conclusion and Future Work}

In this paper, we present \dataset, a comprehensive and high-quality benchmark for C/C++ defects that significantly advances the evaluation and fine-tuning of LLM-based automated program repair techniques. Our dataset fills a critical gap in the field by offering a large-scale, highly usable resource specifically tailored to C/C++ faults. Through extensive experiments on both pre-trained and fine-tuned models, we uncover several important findings. In particular, our evaluation of pre-trained LLMs reveals a substantial performance gap when addressing C/C++ defects compared to their performance on Java-based benchmarks such as Defects4J. This highlights the need for further research on C/C++ program repair.



\section{Acknowledgements}

This work was partially supported by the National Research Foundation, Singapore, and the Cyber Security Agency under its National Cybersecurity R\&D Programme (CRPO-GC1-NUS-001), the CyberSG R\&D Cyber Research Programme Office, the Singapore Ministry of Education Academic Research Fund Tier 1 (RG12/23). Any opinions, findings, and conclusions or recommendations expressed in this material are those of the author(s) and do not necessarily reflect the views of National Research Foundation, Singapore, Cyber Security Agency of Singapore, CyberSG R\&D Programme Office as well as MOE.








\bibliographystyle{IEEEtran}
\bibliography{ijcai25}









\end{document}